\documentclass[prl,a4paper,aps,twocolumn,nofootinbib,nobibnotes,superscriptaddress,preprintnumbers]{revtex4}

\pdfoutput=1

\usepackage[dvips]{graphicx}
\usepackage{amsmath,amssymb,mathrsfs}
\usepackage{bm}
\usepackage{times}
\usepackage{epsfig}
\usepackage{verbatim}
\usepackage{bm}
\usepackage[utf8]{inputenc}
\usepackage{graphics}
\usepackage{graphicx,epsfig,amssymb,amsmath,color,cancel}
\usepackage{subfigure}
\usepackage[english]{babel}
\usepackage{soul}

\begin{document}

\preprint{DESY 18-029}

\title{The Baryon Asymmetry from a Composite Higgs}

\author{Sebastian Bruggisser}
\author{Benedict von Harling}
\author{Oleksii Matsedonskyi}

\affiliation{DESY, Notkestra{\ss}e 85, D-22607 Hamburg, Germany}
\author{G\'eraldine Servant}
\affiliation{DESY, Notkestra{\ss}e 85, D-22607 Hamburg, Germany}
\affiliation{II. Institute of Theoretical Physics, University of Hamburg, D-22761 Hamburg, Germany}

\date{\today}

\begin{abstract}

We study the nature of the electroweak phase transition (EWPT) in models where the Higgs emerges as a pseudo-Nambu-Goldstone boson of an approximate global symmetry of a new strongly-interacting sector confining around the TeV scale.    
Our analysis focusses for the first time on the case where the EWPT is accompanied by the confinement phase transition of the strong sector. 
We describe the confinement in terms of the dilaton, the pseudo-Nambu-Goldstone boson of spontaneously broken conformal invariance of the strong sector. 
The dilaton can either be a meson-like or a glueball-like state and we demonstrate a significant qualitative difference in their dynamics.
We show that the EWPT can naturally be strongly first-order, due to the nearly-conformal nature of the dilaton potential. 
Furthermore, we examine the sizeable scale variation of the Higgs potential parameters during the EWPT.
 In particular, we consider in detail the case of a varying top quark Yukawa coupling, and show that the resulting CP violation 
is sufficient for successful electroweak baryogenesis. We demonstrate that this source of CP violation is compatible with existing flavour and CP constraints.
Our scenario can be tested in complementary ways: by measuring the CP-odd top Yukawa coupling in electron EDM experiments, by searching for dilaton production and deviations in Higgs couplings at colliders,  and through gravitational waves at LISA.

\end{abstract}

\maketitle

\section{Introduction}

Deciphering the origin of the Higgs  potential and its stabilization  against quantum corrections is an essential step towards the microscopic understanding of electroweak (EW) symmetry breaking. 
 One of very few known options for a natural underlying dynamics is that the Higgs boson is a composite object, a bound state of a new strongly interacting sector which confines around the TeV scale~\cite{Panico:2015jxa}. The mass gap between the Higgs and the yet unobserved other composite resonances can be  explained if the Higgs is 
a pseudo-Nambu-Goldstone boson of a global symmetry
$G$ of the strong sector which breaks down to a subgroup $H$ due to a strong condensate $\chi$. The Higgs mass is then protected by a shift symmetry.

Another question left unanswered by the Standard Model (SM) is the origin of the matter-antimatter asymmetry of the universe. One fascinating framework, the EW baryogenesis mechanism~\cite{Morrissey:2012db,Konstandin:2013caa}, 
fails  in the SM due to the absence of a first-order EW phase transition (EWPT) and of sufficient CP-violation.
Determining the nature of the EWPT is an indispensable step to investigate whether EW baryogenesis is the correct explanation for the baryon asymmetry of the universe.

In Composite Higgs (CH) models, since the Higgs arises only
when a non-zero condensate $\chi$ forms, 
the confinement phase transition and the EWPT are closely linked. 
Nevertheless, so far, studies of the EWPT in CH models considered them separately. They 
either focussed on the confinement phase transition, relying on a 5D 
description~\cite{Creminelli:2001th,Randall:2006py,Nardini:2007me,Hassanain:2007js,Konstandin:2010cd,Konstandin:2011dr,Bunk:2017fic,Dillon:2017ctw,vonHarling:2017yew}, or
assumed that the EWPT takes place after confinement of the strong sector~\cite{Delaunay:2007wb,Grinstein:2008qi,Espinosa:2011eu,Chala:2016ykx}.
The novelty of our work is to consider the  interlinked dynamics between the Higgs and the condensate during the EWPT. We present a detailed analysis of the EWPT associated with the confinement phase transition, 
within a purely four-dimensional framework, 
and show that often both phase transitions happen simultaneously. 
We obtain a strong first-order EWPT, thus solving the first problem of EW baryogenesis in the SM.
Complementing previous studies based on 5D-dual models in which the condensate is a glueball, we  also treat the meson case (motivated by lattice studies~\cite{Aoki:2014oha,Appelquist:2016viq}).

An additional attractive feature of CH models is the explanation of the hierarchy of SM Yukawa couplings as originating from the mixing between elementary and composite fermions~\cite{Kaplan:1991dc,Contino:2006nn}. The resulting Yukawa couplings effectively depend on the confinement scale and are therefore expected to vary during the phase transition. CH models thus automatically incorporate the possibility of varying Yukawa couplings during the EWPT, which was shown to bring sufficient CP violation for EW baryogenesis~\cite{Bruggisser:2017lhc,Servant:2018xcs}. 
Furthermore, the Higgs potential in CH models is intimately tied to the top quark  Yukawa coupling.
Its variation then leads to a large variation of the Higgs potential, making the coupled Higgs-$\chi$ dynamics non-trivial. 
We show that sufficient CP violation is naturally induced from the varying top Yukawa, 
thus solving the second problem of EW baryogenesis in the SM. 
We therefore demonstrate that CH models can naturally give rise to EW baryogenesis, 
even  Minimal Composite Higgs Models~\cite{Agashe:2004rs}.

\section{Higgs + Dilaton Phase Transition }

The Higgs potential at present times
can  be parametrised as a sum of trigonometric functions of $h$~\cite{Panico:2012uw},
\begin{equation}\label{eq:vh0}
V^0 [h]\, = \, \alpha^0 \sin^2\left(\frac{h}{f}\right) \, + \, \beta^0 \sin^4\left(\frac{h}{f}\right),
\end{equation}
where $\alpha^0$ and $\beta^0$ are generated by sources which explicitly break $G$ and are fixed to reproduce the mass and vacuum expectation value ({\it vev}) of the Higgs.  
The scale $f$, balancing the Higgs field in the trigonometric functions, is generated by the strong sector condensate. The currently preferred value is around $f = 0.8$~TeV~\cite{Grojean:2013qca} which we will use in the following.  
{The novel aspect of our work is to promote $f$ to a dynamical field. Generally, one expects the confined theory to feature various interconnected condensates, which in particular break the symmetry $G$ (analogous to the chiral symmetry in QCD) with strength given by $f$. Not all of this complex dynamics is necessarily relevant. Flavour physics motivates the strong sector to be nearly conformal above the TeV scale~\cite{Contino:2010rs}. Confinement is then associated with the spontaneous breaking of conformal invariance. This gives rise to a pseudo-Nambu-Goldstone boson, the dilaton, which we denote as $\chi$. As motivated in Ref.~\cite{cpr,Coradeschi:2013gda,Bellazzini:2013fga,Chacko:2012sy,Megias:2014iwa,Megias:2016jcw}, once the explicit breaking of conformal invariance is sufficiently small, the dilaton can be significantly lighter than the confinement scale. Its lightness and the fact that its \emph{vev} sets all scales in the strong sector then allows to integrate out other dynamical fields (whose values now become a function of $\chi$) and to describe the confinement phase transition in terms of $\chi$ getting a \emph{vev}. In particular, this links $f$ to $\chi$.}
We derive the joint potential for the Higgs and the dilaton.%
The potential~(\ref{eq:vh0}) is minimised at $h_0^2  \simeq -(1/2) (\alpha^0/\beta^0) f^2$.
This suggests that the cosmological evolution of the Higgs and the dilaton are tied to each other, {and we show under which conditions both fields obtain a {\it vev} simultaneously.}
% In particular, the strengths of the EWPT and the confinement phase transition then become linked. 
% Since the latter is governed by the almost conformal dilaton potential, this can naturally lead to a strong first-order %EWPT, i.e.~$h/T>1$ at the transition temperature, 
%as confirmed in the following.

We describe the coupled dynamics of the Higgs and the dilaton by using a large-$N$ expansion for the underlying strongly-coupled gauge theory~\cite{Witten:1979kh}, where $N$ represents the number of colors. 
Each insertion of $\chi$ or $h$ is accompanied by a coupling $g_\chi$ or $g_*$, respectively. By large-$N$ counting, these couplings scale as $\sim 1/\sqrt{N}$ for mesons and $\sim 1/N$ for glueballs of the gauge theory. The Higgs is expected to be a meson in analogy with QCD pions while for the dilaton both meson and glueball cases are possible. Requiring a fully strongly interacting theory in the limit $N\rightarrow 1$, this gives~\cite{Panico:2015jxa}
\begin{equation}
	g_*= g_\chi^{(\text{meson})} = 4 \pi/\sqrt N, \qquad g_\chi^{(\text{glueball})} = 4 \pi/N .
\end{equation} 
The trigonometric functions in $V^0 [h]$ can be represented as power series in $h/f$. Using the large-$N$ scaling together with dimensional analysis, one finds that this has to correspond to a power series in $g_* h/(g_\chi \chi_0)$, where $\chi_0$ is the dilaton {\it vev} today.
This fixes the relation between $f$ and $\chi_0$ as $g_* f = g_\chi \chi_0$.

To account for the variation of the scale balancing $h$ in Eq.~(\ref{eq:vh0}) when $\chi$ varies, the kinetic terms are fixed by dimensional analysis as
\begin{equation}\label{eq:lkin}
{\cal L}_{\text{kin}} = \frac 1 2 (\chi/\chi_0)^2(\partial_\mu h)^2 + \frac 1 2 (\partial_\mu \chi)^2.
\end{equation}

We next turn to the Higgs-independent dilaton potential. 
In an exactly conformal theory, only a term $\chi^4$ can appear which does not allow for a minimum $\chi_0\neq 0$. We therefore break conformal invariance explicitly in the UV by a term $\epsilon \mathcal{O}$ in the Lagrangian, where $\mathcal{O}$ is an operator with scaling dimension $4 + \gamma_\epsilon$. If $0> \gamma_\epsilon \gg -1$, the coefficient $\epsilon$ slowly grows when running from the UV scale down to lower energies until it triggers conformal-symmetry breaking and confinement.
This is reflected by an additional term in the dilaton potential (see e.g.~\cite{Megias:2014iwa})
\begin{equation}
\label{eq:Vchipotential}
V_\chi [\chi] \, = \, c_{\chi} g_\chi^2 \chi^4 \, - \, \epsilon[\chi] \chi^4 
\end{equation}
which allows for a minimum at $\chi_0 \neq 0$. Here the function $\epsilon[\chi]$ is governed by an RG equation with $\beta$-function $\beta\simeq \gamma_\epsilon \epsilon + c_\epsilon \epsilon^2/g_\chi^2$ and
$c_\chi$ and $c_\epsilon$ are order-one coefficients. We will trade $\gamma_\epsilon$ for the dilaton mass $m_\chi$ and fix the remaining constants as $c_\epsilon=0.1$, and $c_{\chi}=0.5$ not far from a naive order-one estimate.

Temperature corrections provide a potential barrier 
(which the potential~(\ref{eq:Vchipotential}) does not feature)
necessary for a first-order phase transition.
Indeed, by dimensional analysis and large-$N$ counting, the free energy of the deconfined phase is given by~\cite{Creminelli:2001th,Randall:2006py,Nardini:2007me}
\begin{equation}
\label{FreeEnergy}
\Delta V_T[\chi=0] \, \sim \, - c  N^2 T^4 \ .
\end{equation}
We choose $c=\pi^2/8$, a value corresponding to ${\cal N}=4 $ $SU(N)$ super-Yang-Mills that is representative of a realistic conformal sector. This is modelled by including the standard one-loop thermal corrections from  $45 N^2/4$  strongly coupled degrees of freedom with mass $m = g_\chi \chi$~\cite{Randall:2006py}.
As the temperature drops, $\chi$ eventually tunnels from 0 to the global minimum at $\chi\simeq\chi_0$ corresponding to a confined phase.

Altogether, the potential of our model reads
\begin{equation}
\label{eq:totalpotential}
V_{\rm tot}[h,\chi] \, = \, (\chi/\chi_0)^4 V_h^0[h] \, + \,  V_{\chi}[\chi] \, + \, \Delta V^{\text{1-loop}}_T[h,\chi]\, ,
\end{equation}
where the prefactor $\chi^4$ indicates that the dilaton {\it vev} is the only source of mass in the theory. Furthermore,  
$\Delta V^{\text{1-loop}}_T$ includes the one-loop thermal corrections from SM particles, the Higgs and dilaton as well as the states reproducing the free energy \eqref{FreeEnergy}.

We have calculated  the tunnelling trajectory and action 
for $O(3)$ and $O(4)$-symmetric bubbles in the two-dimensional field space ($h,\chi$). The phase transition happens at a temperature $T_n$ for which the bubble euclidean action is $S_E \approx 140$.
In Fig.~\ref{fig:trajectory}, we show examples of tunneling trajectories {in the meson case}.
The strength of the phase transition  $h[T_n]/T_n$, where $h[T_n]$ is at the minimum of the Higgs potential at $T_n$,
{needs to be $\gtrsim 1$}, to ensure that sphalerons do not wash out the generated baryon asymmetry. {After the phase transition, the system reheats to the temperature $T_{\rm rh}=(30 \Delta V_{\rm tot}/(\pi^2 g_{\rm dof}^{\rm SM}))^{1/4}$ with $\Delta V_{\rm tot}$ being the energy difference between the true and false vacuum. 
We therefore also have to impose that $h(T_{\rm rh})/T_{\rm rh}\gtrsim1$. This enforces the light dilaton window.}
In the left panel of Fig.~\ref{fig:fig345}, we show how 
the phase transition generally quickly becomes supercooled with growing $N$ and decreasing dilaton mass, 
as found in previous studies of the confinement phase transition focussing on the glueball, e.g.~\cite{vonHarling:2017yew}.
This effect is {much} stronger for the glueball than for the meson dilaton due to the different $N$-scaling of its couplings. {This disfavours the glueball case as the baryon asymmetry is either washed out or diluted by $(T_{\rm rh}/T_n)^3$ after reheating. We therefore concentrate on the meson case.}

We will find in the next section that  $\alpha^0$ and $\beta^0$ in Eq.~(\ref{eq:vh0}) can significantly depend on $\chi$,
but this has little impact on the size of the tunnelling action. 
However, it strongly affects the  tunnelling direction, which  controls  the size of the CP-violating source that we now discuss.

\section{CP violation from varying top mixing}
\label{CPviolation}

A sufficient amount of CP asymmetry can be generated during the EWPT from the phase variation of the top quark Yukawa coupling~\cite{Bruggisser:2017lhc}.
This CP-violating source was considered previously in non-minimal CH models with additional singlet scalar field~\cite{Espinosa:2011eu}, and in a 5D model~\cite{vonHarling:2016vhf}. 
Here we do not rely on these extra ingredients. In CH models, the fermion masses originate from linear interactions between the elementary fermions $q_i$ and composite sector operators: $y_i  \bar q_i {\cal O}_i$.

The dimensionless coefficients $y_i$ are assumed to be of order one in the UV, where the mixings are generated. They run subject to an RG equation with $\beta$-function $\gamma_i y_i + c_i y_i^3/g^2_*$, where $c_i$ are order-one coefficients and the scaling dimension of the operator ${\cal O}_i$ is given by $5/2 +\gamma_i$. The anomalous dimensions $\gamma_i$ can remain sizeable over a large energy range due to an approximate conformal symmetry (see \emph{e.g.}~\cite{Contino:2010rs}).
The RG evolution stops at the confinement scale $\sim \chi$, where the operators map to composite states. This makes the mixings $y_i$  dependent on $\chi$. 
Integrating out the composite states, one obtains the effective SM Yukawa couplings
\begin{equation}\label{eq:yukawavsmixing}
\lambda_q[\chi] \, \sim \, y_{qL}[\chi] \, y_{qR}[\chi]/g_* \,,
\end{equation}
where $L$ and $R$ denote the mixings of the left- and right-handed elementary fermions, respectively. In this framework, the SM fermion mass hierarchy is then explained by order-one differences in the scaling dimensions of the operators ${\cal O}_i$.  
This also offers a natural way to make the top Yukawa $\lambda_t$ vary during the phase transition, as the condensation scale then changes.

\begin{figure}[t]
\begin{center}
\includegraphics[width=200pt]{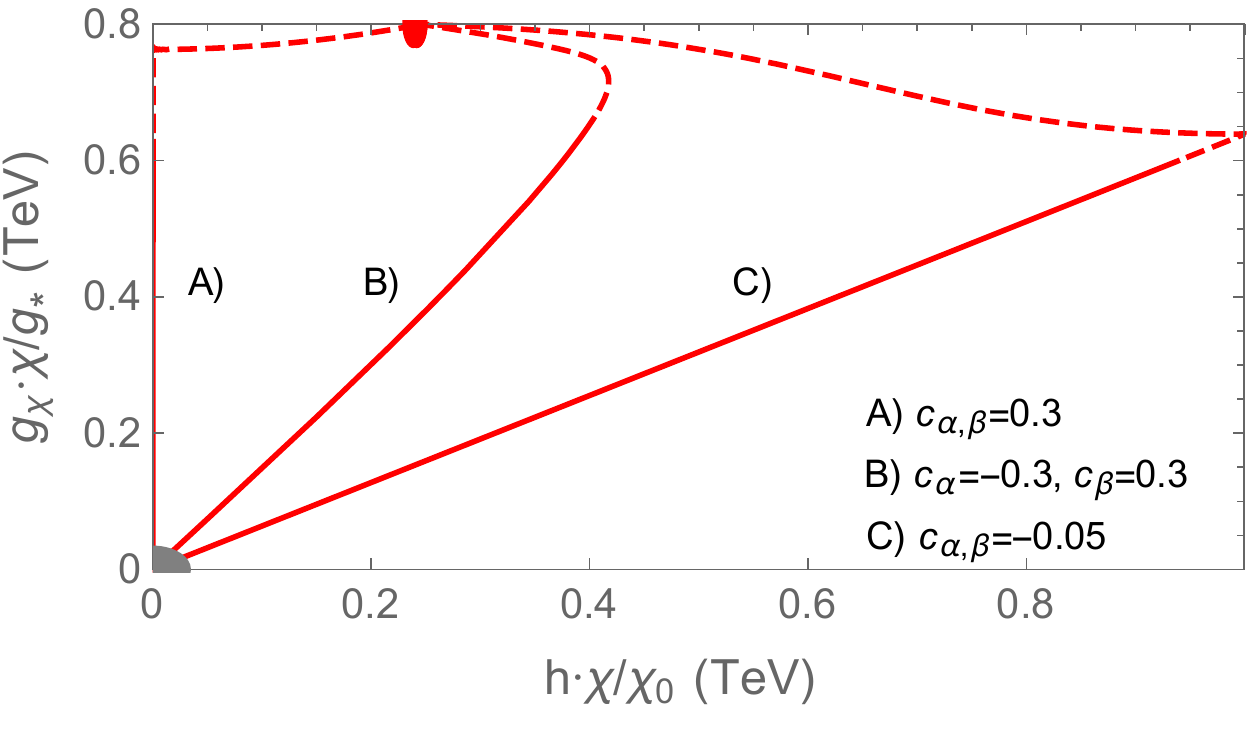}
\end{center}
\caption{\small \it{Transition trajectories for a meson dilaton, $m_{\chi}=700$ GeV, $N=3$.  Solid lines show the tunnelling path to the release point, while dotted lines indicate the subsequent rolling trajectory towards the minimum of the potential at $T_n$, indicated by a bullet.}}
\label{fig:trajectory}
\end{figure}

\begin{figure*}[t]
\begin{center}
%\subfigure{\includegraphics[width=0.32\textwidth]{MNPlotCombined}}\;
%\subfigure{\includegraphics[width=0.32\textwidth]{MNMesonAnglePlot}}\;
%\subfigure{\includegraphics[width=0.34\textwidth]{qqhygammay_1400.pdf}} \hspace{-1.1cm}
\vspace{-0.3cm}
\includegraphics[width=500pt]{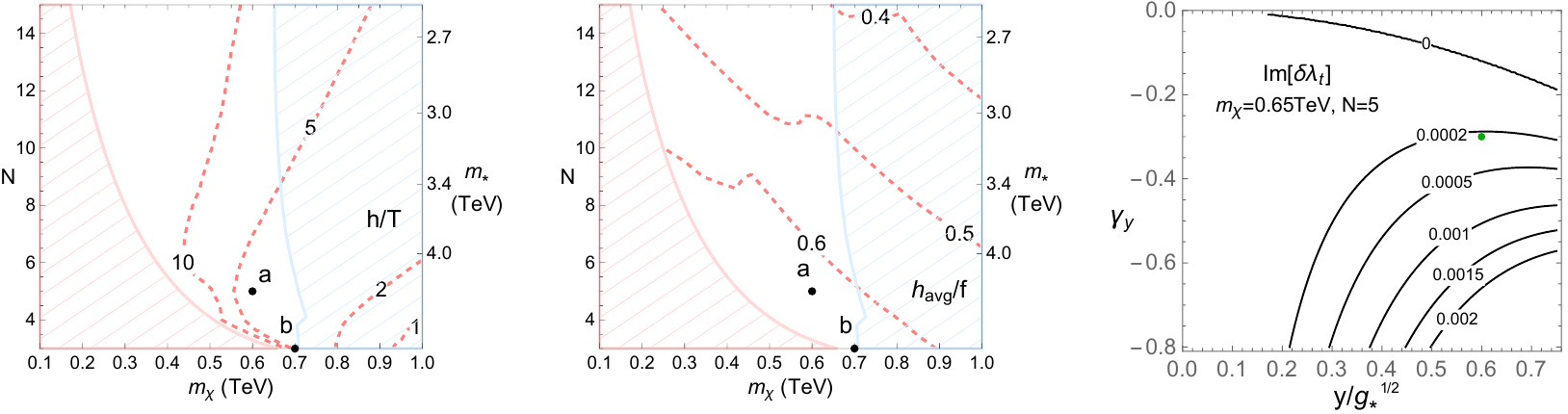}
\end{center}
\caption{\small \it{
%Black solid (red dashed) contours are for a  glueball (meson) dilaton. 
{Results for the meson dilaton.}
In the red dashed region, {no viable EW minimum can be found or the Higgs-dilaton mixing leads to too large deviations in the Higgs couplings.  In the blue dashed region, the baryon asymmetry is washed out after reheating.}  
We also show the cutoff $m_* = g_* f$. The chosen mass range satisfies current experimental constraints~\cite{Blum:2014jca}. Left: Phase-transition strength $h[T_n]/T_n$.
The baryon asymmetry for benchmark point a (b) is $| \eta_B| \times 10^{10}\sim$ 5--5.5 (4--4.5).
Center: Average 
Higgs \emph{vev} during the phase transition relative to the condensate scale today, $h_{\rm avg}/f$.
Right:  
Imaginary part of the top Yukawa as a function of the present value of $y/g_*^{1/2}$ and its anomalous dimension $\gamma_y$ for $|\beta_{y}| = \gamma_y y$,  $\arg \beta_y = 0.1$ and $y_{tL}=\sqrt g_*$. 
The current  and near future experimental sensitivities correspond respectively to approximately $2\times 10^{-2}$~\cite{Cirigliano:2016nyn} and $2\times 10^{-4}$~\cite{Kumar:2013qya}. The green bullet indicates the values used for the left and centre plots.}}
\label{fig:fig345}
\end{figure*}

For the CP-violating source to be non-vanishing, however, $\lambda_t$ needs to vary not only in absolute value but also in phase~\cite{Bruggisser:2017lhc}. To achieve this, we will assume that the right-handed top couples to two different operators in the UV:
\begin{equation}
\label{eq:irpc2}
y_{tR}^{(1)}  \bar t_R {\cal O}_1 +  y_{tR}^{(2)}  \bar t_R {\cal O}_2  \; \, \Rightarrow \; \,  \lambda_t \, \sim \,  y_{tL} (y_{tR}^{(1)}  +  y_{tR}^{(2)})/g_*.
\end{equation} 
Provided that $y_{tR}^{(1,2)}$ are complex and ${\cal O}_{1,2}$ have different scaling dimensions (which we assume to be the case), the phase of $\lambda_t$ changes with $\chi$. 
This provides a source of CP violation, but also has another crucial effect on the phase transition which we now explain. 

The largest contribution to the Higgs potential in CH models typically arises from the top quark mixings. 
We assume that only one of the mixings $\smash{y_{tR}^{(1,2)}}$, which we denote as $y$, varies sizeably with the dilaton {\it vev}. Its one-loop contribution to the coefficients $\alpha^0$ and $\beta^0$ in Eq.~(\ref{eq:vh0}) reads
\begin{equation}
\label{eq:newab}
\alpha[\chi]  =  c_\alpha  \, {3 y^{2}[\chi] g^2_*\over (4 \pi)^2 } f^4 ,\ \  \ 
\beta[\chi]  =  c_\beta  \,  {3  y^{2}[\chi] g^2_*\over (4 \pi)^2 }    f^4 \left({y[\chi]\over g_*}\right)^{p_\beta} ,
\end{equation} 
where $c_\alpha$ and $c_\beta$ are free parameters of our effective field theory, expected to be of order one. Furthermore, $p_\beta=0,2$ depending on the structure of the elementary-composite mixings~\cite{Matsedonskyi:2012ym,Panico:2012uw} (we choose $p_\beta=0$ for definiteness).

%Note that the contributions from the top mixings are somewhat larger than the coefficients $\alpha^0$ and $\beta^0$ %in Eq.~\eqref{eq:vh0} which reproduce the observed Higgs mass and {\it vev}.
%This points to the well-known tuning required to obtain the observed Higgs mass and {\it vev} in CH models: %Additional contributions to $\alpha^0$ and $\beta^0$ must partially cancel those in Eq.~\eqref{eq:newab}.
%However, we can expect this cancellation to happen only when the mixings have their values today and thus, as the %mixings depend on $\chi$, only for $\chi=\chi_0$. 

Notice that this makes the coefficients explicitly depend on $\chi$.
In order to take this into account, we make the 
replacement~\cite{Panico:2012uw}
\begin{equation}
\label{Replacement}
\alpha^0 \to \alpha^0 + (\alpha[\chi] - \alpha[\chi_0]),\quad \beta^0 \to \beta^0 + (\beta[\chi] - \beta[\chi_0])
\end{equation} 
in Eq.~\eqref{eq:vh0}. Furthermore, since the mixings explicitly break the conformal invariance of the CH sector, we include an additional contribution $\propto y^2 \chi^4$ in the dilaton potential (which only plays a subdominant role though).

To have the minimum of the Higgs potential at $h_0\ll f$ at present times requires that $|\alpha^0/\beta^0|\ll1$. From Eq.~(\ref{eq:newab}), on the other hand, we see that generically $|\alpha[\chi]/\beta[\chi]|\gtrsim1$. This is a manifestation of the 
well-known tuning required to obtain the observed Higgs mass and {\it vev} in CH models.

For $\chi$ somewhat away from $\chi_0$, the contributions in Eq.~(\ref{eq:newab}) typically dominate over $\alpha^0$ and $\beta^0$ in Eq.~\eqref{Replacement} and the Higgs potential instead has a global minimum at $h=0$ (for $c_{\alpha,\beta}>0$) or $h=f\pi/2$ (for $c_{\alpha,\beta}<0$). This minimum leads to a valley in the Higgs-dilaton potential which can attract the tunneling trajectory during a first-order phase transition. How closely the tunneling trajectory follows this valley is controlled by its relative depth (in particular  determined by $m_{\chi}$ and $N$) and the value of $\chi$ for which it becomes deeper than the valley along $h=h_0$ that results from the tuned Higgs potential \eqref{eq:vh0} (influenced by $|c_{\alpha,\beta}|, \gamma_y, y[0], y[\chi_0]$). Different tunnelling trajectories are shown in Fig.~\ref{fig:trajectory}. 
The form of the trajectory has major implications for EW baryogenesis. 
In particular, trajectories which closely follow $h=0$ or $h=f\pi/2$ need to be avoided since 
the top mass $\propto \sin[h/f]^{1+m} \cos[h/f]^{n}$~\cite{Pomarol:2012qf} (with $m,n$ being model-dependent)
and thus the CP-violating source vanishes along such trajectories (at $h=f \pi/2$ only if $n\ne 0$). 
%This can also happen for trajectories which closely follow 
%Also for $h=f\pi/2$, 
%however, since the fermion masses are 
%$\propto \sin[h/f]^{1+m} \cos[h/f]^{n}$~\cite{Pomarol:2012qf} with $m,n$ being model-dependent, and therefore %vanish at $h=f \pi/2$ if $n\ne 0$. 

The top mixings are already quite large at ${\chi=\chi_0}$ to ensure a large top Yukawa. Provided that the anomalous dimension $\gamma_y$ for the mixing $y$ is negative, it grows for decreasing $\chi$ until it reaches a fixed point whose size is controlled by the constant $c_y$ in the $\beta$-function.
To obtain a sufficient amount of $y$ variation and CP violation, we choose $\gamma_y=-0.3$ and fix $c_y$ so that $y[0]=0.4 g_*$ in the unbroken phase, while $y[\chi_0]=0.6 \sqrt{\lambda_t g_*}$ in the broken phase. We also set $c_\alpha=-c_\beta=-0.3$ in which case the detuned valley is along $h=f\pi/4$.
We have calculated the action 
%of $O(3)$-symmetric bubbles 
for tunneling along straight lines with constant Higgs \emph{vev} $h$ which well approximates the exact tunneling paths (cf.~Fig.~\ref{fig:trajectory}). In the central panel of Fig.~\ref{fig:fig345}, we plot the Higgs \emph{vev} $h_{\rm avg}$ which minimizes the action at the transition temperature. 
We see that, depending on $m_\chi$ and $N$, different trajectories are possible. 
%Notice that in the case of $h=f \pi /2$ the CP-violating source is non-vanishing only in models with $n=0$.

Thus, as follows from the first two panels in Fig.~\ref{fig:fig345}, the EWPT is strong and our CP-violating source is active for a wide range of $m_\chi$ and $N$. 
We have computed the resulting baryon asymmetry using the formalism presented in Ref.~\cite{Bruggisser:2017lhc}. The results are indicated for {two} benchmark points, assuming a bubble wall velocity of 0.01 (the baryon asymmetry increases by a factor 3-4 if we increase the bubble wall velocity to 0.1) and with the varying mixing in Eq.~\eqref{eq:irpc2} having a complex phase $\arg{y_{tR}^{(1)}}=\arg{y[\chi]}=0.1$ and the remaining mixings being fixed as $ y_{tR}^{(2)}\simeq 0.4\sqrt{\lambda_t g_*}$ and $y_{tL}=\sqrt{\lambda_t g_*}$.

Note that even for $h[T_n]/T_n \gtrsim {\cal O}$(few), we can expect subsonic velocities (needed for baryogenesis) as a sizeable friction comes from the large number of degrees of freedom becoming massive when they go through the bubble wall.
Our baryon asymmetry values (which should only be taken as indicative given order one uncertainties) are typically {close to} the observed value $\eta_B \sim 8.5 \times 10^{-11}$. In contrast with phase transitions studied so far, our Higgs  {\it vev} grows very large during the EWPT before decreasing, and since $\eta_B$ scales as the integral of $(h/T)^2$ over the bubble wall, this leads to a large baryon asymmetry. Furthermore, we find that the bubble wall width $L_w$ is small, also contributing to a large baryon asymmetry.
However, we actually enter a regime where the derivative expansion used in the EW baryogenesis formalism ($L_wT \gg 1$)~\cite{Bruggisser:2017lhc} starts to break down.

\section{Experimental signatures}

The experimental signatures of our scenario include those related to the transition path and CP violation, and those related to the phase-transition strength. The former are strongly sensitive to the $y$ running. 
%Our predictions can be divided into two types - those related to the phase-transition strength (only weakly sensitive to the $y$ running), and those related to the transition path and CP violation (strongly sensitive to the $y$ running). For the former, our testable prediction is the correlation between the dilaton mass and the strong-sector coupling, from the requirement of a strong enough EWPT, see Fig.~\ref{fig:fig345}.
The running mixing $y$ can have a measurable effect on both the Higgs and the dilaton phenomenology, as well as on observables which are indirectly sensitive to the couplings of $h$ and $\chi$. 
Many of these effects arise from the term responsible for the top mass, which in the meson case with $n=0$ reads 
\begin{equation}
\lambda_t[\chi] \, \chi \, \sin\frac h f \, \bar t_L t_R \, \supset \,  
\bar t_L t_R  \, h  \left(\lambda_t^0 \frac{\chi}{f} + \beta_{\lambda_t} \frac{\chi-f}{f}\right), 
\end{equation}  
where $\lambda_t^0$ is the SM top Yukawa coupling, and for one varying mixing we have $\beta_{\lambda_t} \sim \beta_{y}$ (see Eq.~\eqref{eq:yukawavsmixing}). 
$\chi$ and $h$ in this expression are linear combinations of the mass eigenstates.
%, rotated with respect to $h$ and $\chi$ with an angle which also depends on $\beta_{y}$, since $y$ enters into the scalar potential~(\ref{eq:totalpotential}).
Importantly, $\beta_{\lambda_t}$  is complex, as required by the varying  Yukawa phase. The highest sensitivity to the resulting complex couplings comes from measurements of the electron electric dipole moment~\cite{Afach:2015sja}. These restrict the CP-odd coupling of the --mass eigen state-- Higgs to the top (coming from the CP-odd coupling of the --non-mass eigen state-- dilaton) to be
$\lesssim2 \times10^{-2}$ at 95\% CL~\cite{Cirigliano:2016nyn}, with a prospect of gaining about two orders of magnitude  in sensitivity in the near future~\cite{Kumar:2013qya}.  
In the right panel of Fig.~\ref{fig:fig345}, we show how the CP-odd $tth$ coupling depends on $y[\chi]$. 
Forthcoming experiments are expected to probe { most of} our parameter space. 

The strength of the phase transition in our model becomes linked to the dilaton mass, which is light, hence can be searched for in collider experiments. {Higgs-dilaton mixing also leads to observable deviations in the Higgs couplings.}
Another related signature is a stochastic background of gravitational waves peaked in the milli-Hertz range that can be searched for at LISA~\cite{Randall:2006py,Bruggisser:2018mrt}.

% transition strength, instead,  directly related to the phase transition strength are the 
%On the other hand, the experimental te

%In the longer term, future colliders can probe deviations in the Higgs couplings arising from the mixing with the dilaton~\cite{ourlongpaper}  and   a stochastic background of gravitational waves peaked in the milli-Hertz range can be searched for at LISA~\cite{Randall:2006py,ourlongpaper}.

In summary, our results strongly support the viability of EW baryogenesis and motivate further studies in concrete calculable realizations of CH models. In a forthcoming paper \cite{Bruggisser:2018mrt}, we extend this analysis to other possible sources of CP violation.

\bibliography{biblio} 

\end{document}